\documentclass[aps,showpacs,preprintnumbers,amsmath,amssymb,nofootinbibt,preprint]{revtex4}
\usepackage{graphicx}
\usepackage{amsmath}
\usepackage{amssymb}
\usepackage{amsfonts}
\usepackage{bm}
\usepackage{color}

\def\be{\begin{equation}}
\def\ee{\end{equation}}
\def\bea{\begin{eqnarray}}
\def\eea{\end{eqnarray}}

\begin{document}

\title{Gravitational Effects of Condensate Dark Matter on Compact Stellar Objects}
\author{X. Y. Li}
\email{lixinyu@connect.hku.hk}
\affiliation{Department of Physics, The University
of Hong Kong, Hong Kong, China and
Perimeter Institute for Theoretical Physics, Ontario, Canada }
\author{F.Y. Wang}
\affiliation{Department of Physics, The University
of Hong Kong, Hong Kong, China and Department of Astronomy, Nanjing University, Nanjing, China}
\author{K. S. Cheng}
\affiliation{Department of Physics, The University
of Hong Kong, Hong Kong, P. R. China}

\begin{abstract}
We study the gravitational effect of non-self-annihilating dark matter on compact stellar objects. The self-interaction of condensate dark matter can give high accretion rate of dark matter onto stars. Phase transition to condensation state takes place when the dark matter density exceeds the critical value. A compact degenerate dark matter core is developed and alter the structure and stability of the stellar objects. Condensate dark matter admixed neutron stars is studied through the two-fluid TOV equation. The existence of condensate dark matter deforms the mass-radius relation of neutron stars and lower their maximum baryonic masses and radii. The possible effects on the Gamma-ray Burst rate in high redshift are discussed.
\end{abstract}

\maketitle

\section{Introduction}

In the $\Lambda$CDM cosmology model,  dark matter contributes 24\% of total energy of the universe. Dark matter is regarded as the majority of matter in the universe, and it has significance effect on large-scale structure formation and evolution. Astrophysical observations provided indirect means to probe dark matter and study its effects.

Dark matter may also have effects on individual stellar objects. Steigman et al. \cite{st78} put forward the idea that capture of WIMP particle would affect the stellar structure and evolution. Spergel and Press \cite{sp85} and Faulkner and Gilliland \cite{fg85} employed the idea for an explanation of the solar neutrino problem. Recently, there are growing interests in this field.  Self-annihilating dark matter provide extra energy for stellar objects, which have been extensively studied by \cite{bf08}, \cite{k08}, \cite{gn89}, \cite{gd90}. In particular its impact on first-generation stars is studied by \cite{sp09}, \cite{spap09}, \cite{rip10}. The impact on the evolution path of main sequence star is studied by \cite{cl09}, \cite{cl11}.  For non self-annihilating dark matter, its impact is studied by \cite{fs10}, \cite{cg10}, \cite{ti10} for main sequence stars and for neutron stars by \cite{cs11},  \cite{sc09}, \cite{lcl11} \cite{bh07} for different dark matter models.

In this paper, we study the gravitational effect of non-self-annihilating condensate dark matter  on stellar objects. The Bose-Einstein condensation may come from the bosonic features of dark matter models, e.g. axions, Nambu-Goldston bosons or supersymmetric partner of fermions suggested by \cite{sin1994},\cite{overduin2004},\cite{hu2000}. Some previous studies of boson star structure and dynamics can be found in \cite{ss96}, \cite{bs98}, \cite{bc98},\cite{ch2012}. Condensate dark matter has an equation of state $P_{\chi}=\frac{2\pi \hbar^{2}l_{a}}{m_{\chi}^{3}}\rho_{\chi}^{2}$, where $l_{a}$ is the scattering length and $m_{\chi}$ is the dark matter particle mass. Condensate dark matter is able to form gravitationally bound objects after the condensation takes place. The properties, structure and stability of such objects is studied by \cite{li12}. Phase transition to condensation can occur under either one of the following two conditions: when the temperature cools below critical value or when the density exceeds the critical value \cite{li12}. In this paper we consider the following picture: the deep gravitational well of a star accretes dark matter, when phase transition takes place, dark matter forms a compact degenerate dark matter core inside the star and may causee the star to collapse to compact object, i.e. white dwarf, neutron star or black hole. We examine conditions of the existence of such degenerate core can make the white dwarf or neutron star stable. If no stable neutron star is possible, the star will eventually collapse to a black hole. The accretion rate and criteria for phase transition are presented in this paper. Following that, analytical and numerical study for the structure and stability of various stellar objects with condensate dark matter component are presented. The parameters for condensate dark matter are $l_{a}=1\rm fm$, $m_{\chi}=1\rm GeV$ as fiducial values.

\section{Accretion of Dark Matter onto Stars}\label{accr}
The self-interaction of condensate dark matter enables the dark matter particles to transfer energy to each other. Through interacting and losing energy to dark matter particles already inside the star, incoming dark matter particles will be captured by the star and stay inside it. The calculation of accretion rate of dark matter particle follows from Gould \cite{gou87}, while self-interaction is considered the interaction between dark matter and baryons is ignored. The self-interaction cross section of dark matter is determined by
\begin{equation}
\sigma_{\chi\chi}=4\pi l_{a}^{2}.
\end{equation}

Assuming spherically symmetric gravitational field for stars, the escaping velocity is $v$. Dark matter has velocity distribution $f(u)\mathrm{d} u$. $\Omega_{v}^{-}(w)$ is defined as the rate per unit time that a dark matter particle with velocity $w$ will be scattered to a velocity less than $v$ when collided with materials. The flux of dark matter particles that goes inward is
\begin{equation}\label{3.1}
\frac{1}{4}f(u)u\mathrm{d} u\mathrm{d} \cos^{2}\theta,
\end{equation}
where $\theta$ is the angle relative to the radial direction, $0\leq\theta\leq\frac{\pi}{2}$. The probability for a dark matter particle to be scattered to velocity less than $v$ is
\begin{equation}\label{3.5}
\Omega_{v}^{-}(w)\frac{\mathrm{d} l}{w},
\end{equation}
where
\begin{equation}\label{3.6}
\frac{\mathrm{d} l}{w}=\frac{2}{w}\Big[1-\Big(\frac{J}{rw}\Big)^{2}\Big]^{-\frac{1}{2}}\mathrm{d} r \Theta(rw-J).
\end{equation}
$J$ is the angular momentum of an incoming dark matter particle. $\Theta(rw-J)$ is the step function. After integration over whole angular momentum space, the total capture rate per unit shell volume is
\begin{equation}
\frac{\mathrm{d} C}{\mathrm{d} V}=\int_{0}^{\infty}\mathrm{d} u\frac{f(u)}{u}w\Omega_{v}^{-}(w).
\end{equation}
Let the dark matter particles with mass $m_{\chi}$ and number density $n^{\rm in}$ in the star. From kinematics, the energy loss in a collision, $\frac{\Delta E}{E}$ lies in the interval
\begin{equation}
0\leq\frac{\Delta E}{E}\leq1,
\end{equation}
The net capture of a dark matter particle requires
\begin{equation}
\frac{w^{2}-v^{2}}{w^{2}}=\frac{u^{2}}{w^{2}}\leq\frac{\Delta E}{E}\leq\frac{v^{2}}{w^{2}}.
\end{equation}
The total probability of capturing a dark matter particle is
\begin{equation}
\Omega_{v}^{-}(w)=\sigma_{\chi\chi} n^{\rm in}w\frac{v^{2}-u^{2}}{w^{2}}\Theta\left(v^{2}-u^{2}\right).
\end{equation}
Taking the incoming dark matter particle follow the Maxwell-Boltzmann distribution with number density $n^{\rm out}$,
\begin{equation}
f(u)\mathrm{d} u=n^{\rm out}\frac{4}{\sqrt{\pi}}x^{2}\exp(-x^{2})\mathrm{d} x,
\end{equation}
where $x$ is the dimensionless velocity defined by
\begin{equation}
x^2=\frac{m_{\chi}u^{2}}{2k_{B}T_{\chi}}\equiv\frac{3u^{2}}{2\overline{v}^{2}}
\end{equation}
with $\overline{v}$ being the average velocity of dark matter particles in galaxies.

The capture rate per unit shell volume is
\begin{equation}
\frac{\mathrm{d} C}{\mathrm{d} V}(r)=\left(\frac{6}{\pi}\right)^{\frac{1}{2}}\sigma_{\chi\chi} n^{\rm in} n^{\rm out}\frac{v^{2}}{\overline{v}}\left[1-A^{-1}(1-\exp(-A))\right],
\end{equation}
where $A=\frac{3}{2}\frac{v^{2}}{\overline{v}^{2}}$. Generally $A\gg1$, the mass accretion rate is given by
\begin{equation}
\dot{M_{\chi}}=\int m_{\chi}\mathrm{d} C=\sqrt{\frac{6}{\pi}}M_{\chi}(t)\sigma_{\chi\chi}n^{\rm out}\frac{v^{2}}{\overline{v}}.
\end{equation}
So the total mass accreted at time $t$ is
\begin{equation}
M_{\chi}(t)=M_{0}\exp(\sqrt{\frac{6}{\pi}}\sigma_{\chi\chi}n^{\rm out}\frac{v^{2}}{\overline{v}}).
\end{equation}

The accreted dark matter mass grows exponentially. However, it is only valid when the probability term (\ref{3.5}) is smaller than 1 and shall never exceeds 1. For a star with radius $R_{*}$, an estimation of when the probability reaches 1 can be given by
\begin{equation}
n^{\rm in}\sigma_{\chi\chi}R_{*}=1.
\end{equation}
When the probability term (\ref{3.5}) equals 1, all incoming dark matter is accreted onto the star. So the capture rate is simply integration of (\ref{3.1}) over the star surface.
\begin{eqnarray}
\dot{M_{\chi}}&=&\pi R_{*}^{2}\int_{0}^{\infty}m_{\chi}f(u)u\mathrm{d} u=2\sqrt{\frac{2\pi}{3}}\mathit{R_{*}^{2}}n^{\rm out}\overline{v}m_{\chi}\nonumber\\
&=&2\times10^{-10}M_{\odot}/\rm yr\left(\frac{\mathit{R_{*}}}{10^{11}\rm cm}\right)^{2}\left(\frac{\mathit{\overline{v}}}{250\rm km/s}\right)\left(\frac{\rho^{\rm out}}{10^{10}\rm GeV/cm^{3}}\right).
\end{eqnarray}
Hence, the total accretion rate is given by
\begin{displaymath}
\dot{M_{\chi}}=\left\{\begin{array}{ll}
\sqrt{\frac{6}{\pi}}M_{\chi}\sigma_{\chi\chi}n^{\rm out}\frac{v^{2}}{\overline{v}}\; &n^{\rm in}\sigma_{\chi\chi}R_{*}<1,\\
2\sqrt{\frac{2\pi}{3}}R_{*}^{2}n^{\rm out}\overline{v}m_{\chi}\; & n^{\rm in}\sigma_{\chi\chi}R_{*}\geq1.
\end{array}\right.
\end{displaymath}
In our calculation, dark matter mass will first grow exponentially. After reaching a certain mass, the accretion rate becomes constant.

The total dark matter mass accreted is given by
\begin{displaymath}
M_{\chi}(t)=\left\{\begin{array}{ll}
M_{\chi0}\exp\left[5.83\left(\frac{l_{a}}{1\rm fm}\right)^{2}\left(\frac{n^{\rm out}}{10^{10}}\right)\left(\frac{v}{516\rm km/s}\right)^{2}\left(\frac{\overline{v}}{250\rm km/s}\right)^{-1}\left(\frac{t}{1\rm yr}\right)\right]\; &n^{\rm in}\sigma_{\chi\chi}R_{*}<1,\\
M_{\chi1}+2\times10^{-10}M_{\odot}\left(\frac{\mathit{R_{*}}}{10^{11}\rm cm}\right)^{2}\left(\frac{\mathit{\overline{v}}}{250\rm km/s}\right)\left(\frac{n^{\rm out}}{10^{10}}\right)\left(\frac{t}{1\rm yr}\right)\; & n^{\rm in}\sigma_{\chi\chi}R_{*}\geq1.
\end{array}\right.
\end{displaymath}
$M_{\chi0}$ is the initial mass of dark matter inside the star before accretion and $M_{\chi1}$ is the mass of dark matter inside the star when the probability term (\ref{3.5}) reaches 1. With typical values of parameters of condensate dark matter, the constant accretion rate is reached very quickly.  For a  star with solar mass and radius $10^{11}\rm cm$, let $n^{\rm out}=10^{10}$, $\frac{3M_{\chi0}}{4\pi R^{3}}=m_{\chi}n^{\rm out}$, the probability takes place when $n^{\rm in}\approx10^{14}$. So the exponential accretion lasts for $\frac{\log 10^{4}}{5.83}\approx1.5\rm yr$. The lasting time of exponential accretion is inverse proportional to the logarithm of initial mass, which will be very small compared to the stellar lifetime. Therefore the constant accretion stage actually determines that the amount of dark matter can be accreted within the stellar lifetime.  The total dark matter accreted within the stellar life is proportional to the ambient dark matter density, if the intrinsic properties of dark matter are fixed.

The density of incoming dark matter particles near the star constrains the accretion rate. In the early universe the ambient dark matter density is much higher than today. When the early stars are formed, the creation of deep gravitational well in the dark matter halo makes it to further contract and increase the density of dark matter near stars to be even higher.

The increase of dark matter density in star formation can be calculated by using the adiabatic approximation. The adiabatic invariant is $rM(r)$ where $M(r)$ denotes the total mass enclosed within radius $r$. This leads to the adiabatic equation given by Blumenthal et al. \cite{blu86}
\begin{equation}\label{ac}
M(r_{i})r_{i}=\frac{M_{\chi}(r_{i})}{1-f_{b}}=(M_{\chi}(r_{i})+M_{b}(r_{f}))r_{f}.
\end{equation}
$M(r)$, $M_{\chi}(r)$ and $M_{b}(r)$ are respectively the total mass, dark matter mass and baryon mass inside radius $r$. $r_{i},r_{f}$ denote the radius before and after the contraction inside which dark matter of total mass $M_{\chi}(r_{i})$ is enclosed, $f_{b}=0.15$ is the mass fraction of baryon in the initial dark matter halo.

\begin{figure}[htbp] %  figure placement: here, top, bottom, or page
   \centering
   \includegraphics[width=4in]{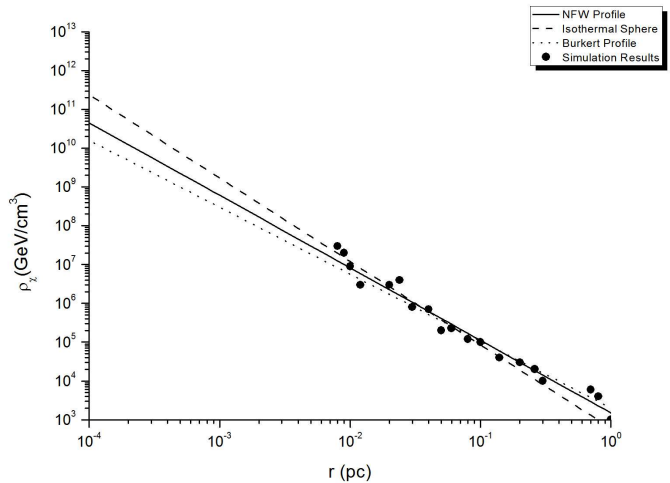}
   \caption{Comparison Between Adiabatic Contracted Density Profile of Different Initial Dark Matter Distribution and Simulation Result by Abel et al.\cite{abel02}: Dots are simulation results.}
   \label{fig1}
\end{figure}

Fig.\ref{fig1} illustrates the contracted dark matter profile for different initial profiles: NFW profile \cite{nfw96}, isothermal sphere \cite{bt94} and Burkert profile \cite{bkt}. Dots are from the simulation results by Abel et al. \cite{abel02} The parameters of each initial profile are tuned to fit the simulation results. The best fit line is from the NFW profile. The parameter for the dark matter halo is chosen to have total mass $7\times10^{5}M_{\odot}$ and the concentration parameter as in \cite{nfw96} $c=2$. This result agrees with the calculation done by Freese et al. \cite{fr09}

The adiabatic contracted dark matter density profile fits the simulation result quite well from 0.01pc outward, but how close to the core this profile is still valid is unclear, as near the core the strong diffusive process in star formation may break the adiabatic condition of dark matter.

\section{Phase transition of dark matter inside the star}
Condensate dark matter particles will undergo Bose-Einstein condensation inside the star when its density is above the critical one. The critical density for this kind of phase transition is given by \cite{li12}
\begin{equation}
\rho _{\chi,\rm cr}=\frac{\sigma_{v} ^2m_{\chi }^3}{2\pi \hbar ^2 l_a}=7.327\times 10^{9}\left(\frac{\sigma_{v} ^2}{9\times10^{14}\rm cm^{2}/s^{2}}\right)
\left(\frac{m_{\chi }}{1\;{\rm GeV}}\right)^3\left(\frac{l_a}{1\;{\rm fm}}\right)^{-1}\;{\rm g/cm^3},
\end{equation}
where $\sigma_{v}^{2}$ is the velocity dispersion of normal dark matter.

The distribution of dark matter particle inside the star is beyond current knowledge, while it might be possible to extrapolate some from the distribution without the gravitational potential. Simulations and self-similar solutions give that dark matter density of dark matter profile will follow the power-law distribution $\rho\propto r^{-\alpha}$ (cf \cite{nfw96}, \cite{fm97}, \cite{ber85}, \cite{wan00}). The exact value of $\alpha$ is still unclear, but most studies show the value lies in the range $1\leq\alpha\leq3$. With the existence of extra gravitational well due to the baryonic star, the dark matter profile will have a steeper profile, for more dark matter will be dragged into the core by the extra gravity.

In a very small central region that dark matter density may exceed the critical density, the dark matter within this region will become condensate but they can only form some microscopic droplets before they reach the minimum mass of the gravitationally bounded object (cf \cite{li12}), whose exact size depends on the detail of the exact interaction of dark matter and their surface energy.  When dark matter continue to accrete into the core of the star, the number of microscopic droplets increase. When the total mass of these droplets is sufficiently large, they become the gravitationally-bound object. The minimal mass of stable condensate dark matter object is obtained from \cite{li12} by setting the central density to be the lowest one which is the critical density of phase transition. With our selection of dark matter parameters, the minimal mass is $M_{\chi, \rm min}=2.296\times10^{-5}M_{\odot}$.

Taking the total dark matter mass inside the star to be $M_{\chi}(t)=\int \dot{M_\chi}\mathrm{d}t$,  $R_{*}$ to be the stellar mass, and the dark matter density profile to follow
\begin{equation}
\rho_{\chi}=\rho_{\chi,0}\left(\frac{r}{R_{*}}\right)^{-\alpha},
\end{equation}
the dark matter mass in the region where density is above critical value is given by
\begin{equation}
M_{\chi, \rm cen}=M_{\chi}\left(\frac{\rho_{\chi,0}}{\rho_{\chi,\rm cr}}\right)^{\frac{3-\alpha}{\alpha}}.
\end{equation}
The criteria for the condensate dark matter core to form is
\begin{equation}
M_{\chi, \rm cen}>M_{\chi, \rm min}.
\end{equation}

\section{formulation for stars with dark matter components}\label{ana}
We assume that dark matter only interact with baryons through gravity and study the static structure of such stars with dark matter components. We adopt the two-fluid formulation by \cite{cs11}, which is first introduced by \cite{sc09}. This formulation is a separation of TOV equation motivated by the similarity of structure equations between the relativistic and Newtonian ones.  There are two equations for the balance between gravity and pressure for baryons and dark matter separately.
\begin{eqnarray}
r^{2}\frac{\mathrm{d}P_{\rm B}(r)}{\mathrm{d}r}&=&-G M(r)\left[1+\frac{4\pi r^{3}P(r)}{M(r)c^{2}}\right]\left[1-\frac{2GM(r)}{c^{2}r}\right]^{-1}\left(\rho_{\rm B}(r)+\frac{P_{\rm B}(r)}{ c^{2}}\right),\nonumber\\
r^{2}\frac{\mathrm{d}P_{\rm DM}(r)}{\mathrm{d}r}&=&-G M(r)\left[1+\frac{4\pi r^{3}P(r)}{M(r)c^{2}}\right]\left[1-\frac{2GM(r)}{c^{2}r}\right]^{-1}\left(\rho_{\rm DM}(r)+\frac{P_{\rm DM}(r)}{ c^{2}}\right),
\end{eqnarray}
where $P(r)=P_{\rm B}(r)+P_{\rm DM}(r)$ and $M(r)=M_{\rm B}(r)+M_{\rm DM}(r)$.
With two more equations of mass continuity of baryon and dark matter separately,
\begin{eqnarray}
\frac{\mathrm{d}M_{\rm B}(r)}{\mathrm{d}r}&=&4\pi r^{2}\rho_{\rm B}(r),\nonumber\\
\frac{\mathrm{d}M_{\rm DM}(r)}{\mathrm{d}r}&=&4\pi r^{2}\rho_{\rm DM}(r),
\end{eqnarray}
the four equations give a complete set of equations for the structure of stars with dark matter component.

Even though there are two-fluid formulation available from first principle calculation \cite{carter}, they give identical numerical results for static stellar cases\cite{chu}.  Therefore, we take this simple approach to study the static stellar structure with dark matter component.

Once dark matter condensation takes place, the gravitationally bound object formed has density at least $7.327\times10^{9}\rm g/cm^{3}$ \cite{li12}, much greater than the onset of electron degeneracy ($10^{4}-10^{5}\rm g/cm^{3}$) \cite{bwn} and central density for most white dwarfs ($10^{4}\rm g/cm^{3}-10^{10} \rm g/cm^{3}$) \cite{ta}. For normal stars or white dwarfs, the baryonic component inside the dark matter core is negligible. The normal dark matter interacting with the condensate core will also rest to the ground state, become condensate and accrete onto the core. Therefore, we assume that all dark matter inside the star are concentrated in the degenerate core.  Under this approximation, the star has a pure dark matter core and outside the core is pure baryonic matter. We treat the pure dark matter core as initial conditions for the normal TOV equations and integrate the equations for baryonic matter from the surface of the dark matter core.

The normal TOV equations for stars is given by
\begin{eqnarray}
\frac{\mathrm{d}M}{\mathrm{d}r}&=&4\pi r^{2}\rho(r),\nonumber\\
\frac{\mathrm{d}P}{\mathrm{d}r}&=&-\frac{G}{r^{2}} \left[M(r)+\frac{4\pi r^{3}P}{c^{2}}\right]\left[1-\frac{2GM(r)}{c^{2}r}\right]^{-1}\left(\rho+\frac{P}{ c^{2}}\right).
\end{eqnarray}
The dark matter core has radius $r_{c}$ and mass $M_{c}$ which can be determined from \cite{li12} and the density of normal baryonic matter at the surface of the core is $\rho_{0}$. The initial conditions for the TOV equations for the star with dark matter core are when $r=r_{c}$, $M=M_{c},\; \rho=\rho_{0}$. Using the parametrization below
\begin{eqnarray}
&&\theta=\frac{\rho}{\rho_{0}c^{2}}, \; \sigma=\frac{P}{\rho_{0}c^{2}},\nonumber\\
&&m=M\frac{c^{3}}{G}\sqrt{4\pi G\rho_{0}}, \; \xi=\frac{r}{c}\sqrt{4\pi G\rho_{0}},\nonumber\\
&&m_{c}=M_{c}\frac{c^{3}}{G}\sqrt{4\pi G\rho_{0}}, \; \xi_{c}=\frac{r_{c}}{c}\sqrt{4\pi G\rho_{0}},
\end{eqnarray}
the TOV equations reduce to the following form\cite{fang}
\begin{eqnarray}\label{str}
\frac{\mathrm{d}m}{\mathrm{d}\xi}&=&\xi^{2}\theta^{n},\nonumber\\
\frac{\mathrm{d}\sigma}{\mathrm{d}\xi}&=&-\frac{(\theta+\sigma)(\sigma\xi^{3}+m)}{\xi^{2}(1-\frac{2m}{\xi})},\nonumber\\
\theta(\xi_{c})&=&1,\;m(\xi_{c})=m_{c}.
\end{eqnarray}
When $\xi\rightarrow\xi_{c}$, the equation leads to
\begin{equation}
\frac{\mathrm{d}\sigma}{\mathrm{d}\xi}\rightarrow-\frac{(1+\sigma)(m_{c}+\sigma\xi_{c}^{3})}{\xi_{c}^{2}(1-\frac{2m_{c}}{\xi_{c}})}.
\end{equation}

In order to have physical solution, $\frac{\mathrm{d}\sigma}{\mathrm{d}\xi}<0$ must be satisfied. This constraint leads to
\begin{equation}\label{bh}
2\frac{m_{c}}{\xi_{c}}<1,
\end{equation}
equivalently
\begin{equation}
\frac{2GM_{\rm c}}{c^{2}r_{\rm c}}<1.
\end{equation}
This constraint means that the core cannot directly collapse to a black hole.

Another constraint comes from the non-relativistic limit of Eq.(\ref{str}) in the region where the gravity of dark matter core dominates. So the mass and self-gravitation of normal baryonic matter are neglected, $m\approx m_{c}$. In the non-relativistic limit, $\sigma\ll1$, the TOV equations become
\begin{equation}
-\frac{1}{\theta}\mathrm{d}\sigma=\frac{ m_{c}}{\xi^{2}(1-\frac{2m_{c}}{\xi})}\mathrm{d}\xi.
\end{equation}
Integrating both sides,  since $\theta\leq1$ the left hand side gives
\begin{equation}\label{lhs}
\int_{\xi_{c}}^{\xi}-\frac{1}{\theta}\mathrm{d}\sigma>\sigma(\xi_{c})-\sigma(\xi).
\end{equation}
The right hand side can be integrated analytically,
\begin{equation}
\int_{\xi_{c}}^{\xi}\frac{ m_{c}}{\xi'^{2}(1-\frac{2m_{c}}{\xi'})}\mathrm{d}\xi'=\frac{1}{2}\log\frac{1-\frac{2m_{c}}{\xi}}{1-\frac{2m_{c}}{\xi_{c}}}.
\end{equation}
Since the dark matter core is much smaller than the star, in regions $\xi\gg\xi_{c}$, stable solution requires the pressure to be always positive
\begin{equation}
\sigma(\xi_{c})>-\frac{1}{2}\log\left(1-\frac{2 m_{c}}{\xi_{c}}\right).
\end{equation}
Since in Eq.(\ref{lhs}) the integration takes $\theta=1$, this analysis gives sufficient condition of $\rho_{0}$ to have stable solutions.

In case that baryonic matter has a polytropic equation of state $P=K\rho^{1+1/n}$, Eq.(\ref{lhs}) can be integrated analytically. In addition, if the core potential is non-relativistic $\frac{2GM_{\rm c}}{c^{2}r_{\rm c}}\ll1$, the constraint reduces to the form given by\cite{fang}
\begin{equation}
\frac{K\rho_{0}^{1/n}}{c^{2}}>\frac{GM_{\rm c}}{(1+n)c^{2}r_{\rm c}}.
\end{equation}
This constraint gives the minimum value of $\rho_{0}$ to have stable solutions.
\begin{equation}
\rho_{0,\rm min}=\left[\frac{GM_{\rm c}}{(1+n)Kr_{\rm c}}\right]^{n}.
\end{equation}

\section{Applications}
Since normal stars have very low density, applying the analysis in Section \ref{ana} to them gives that, once the condensate dark matter core is formed, the star is unstable except for extremely light dark matter core. The core of the star is expected to collapse to form compact objects, i.e. white dwarfs or neutron stars. Instead of considering the actually collpase processes we only focus to study the stability of the static structure of white dwarfs and neutron stars with dark matter component.

\subsection{White Dwarfs}
As stated in Section \ref{ana}, the central density of most white dwarfs is much smaller than condensate dark matter and baryon component is negligible inside the dark matter core. We take the condensate dark matter core as initial conditions for the TOV equation and integrate the equation from the core surface. The equation of state (EOS) for baryonic component is taken to be the ideal fermi gas. This EOS is only valid below Neutron drip ($\sim 4\times10^{11}\rm g/cm^{3}$\cite{bwn}) and neutronisation threshold ($\sim1.22\times10^{7}\rm g/cm^{3}$ for inverse beta decay \cite{bwn}). For high central density white dwarfs ($>10^{7}\rm g/cm^{3}$), its equation of state has been calculated by \cite{bbps}. When the central density becomes comparable to that of condensate dark matter, we need to solve the two-fluid equation. The integration of two-fluid equation is incorporated in the next section of neutron star where we use the BBPS EOS for low density region. In this section the upper limit of central density $\rho_{0}$ is set to be $10^{10}\rm g/cm^{3}$ as fiducial value.

\begin{figure}[htbp] %  figure placement: here, top, bottom, or page
   \centering
   \includegraphics[width=4in]{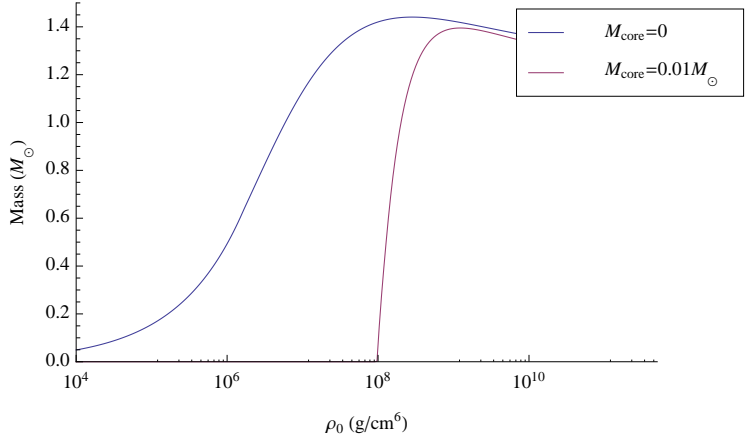}
   \caption{Relation Between Baryon Mass and $\rho_{0}$ for White Dwarf with and without a Dark Matte Core of $0.01M_{\odot}$}
   \label{fig2}
\end{figure}

Fig.\ref{fig2} illustrates the comparison of the relation between total mass (baryon mass plus dark matter mass) and central density for white dwarf with a core of $0.01M_{\odot}$ and that without a core. The existence of the core contributes extra gravity that causes the white dwarf to contract, resulting to a increase of central density and decrease of maximum mass.

\begin{figure}[htbp] %  figure placement: here, top, bottom, or page
   \centering
   \includegraphics[width=4in]{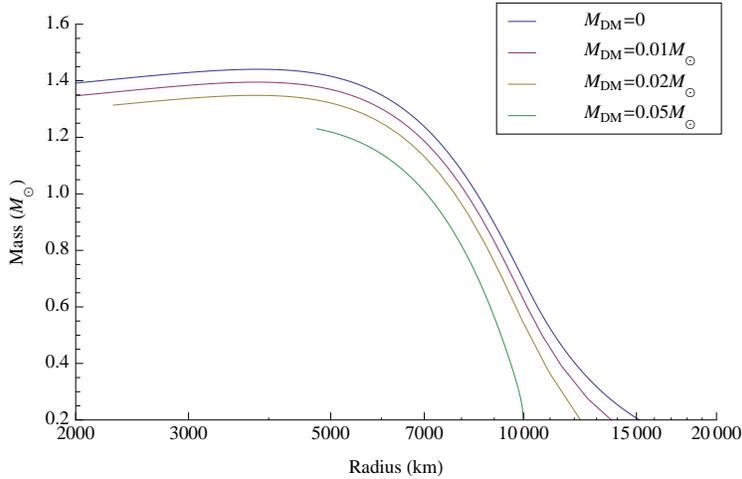}
   \caption{Relation Between Total Mass and Radius  for White Dwarf with Various Dark Matter Core Mass}
   \label{fig3}
\end{figure}

Fig.\ref{fig3} compares the relation between total mass and radius for white dwarfs with various dark matter core mass. The increase of dark matter mass results in both smaller radius and smaller mass. For dark mass smaller than $0.04M_{\odot}$, there is an instability of white dwarf when $\frac{\mathrm{d}M}{\mathrm{d}\rho_{0}}<0$. This criteria gives unstable fundamental mode of radial oscillation \cite{bwn}. However, this elementary stability analysis shall be taken with care, since $\frac{\mathrm{d}M}{\mathrm{d}\rho_{0}}>0$ is not sufficient to conclude the white dwarf is stable. A proper stability analysis shall include all radial eigenmodes which is beyond the scope of this paper. But any white dwarf with higher mass will be definitely unstable.  We determine maximum mass by this instability. When the core mass keeps growing, there is no instability of fundamental radial mode, the maximum mass is determined when the central density of white dwarf reaches maximum $10^{10}\rm g/cm^{3}$.

\begin{figure}[htbp] %  figure placement: here, top, bottom, or page
   \centering
   \includegraphics[width=4in]{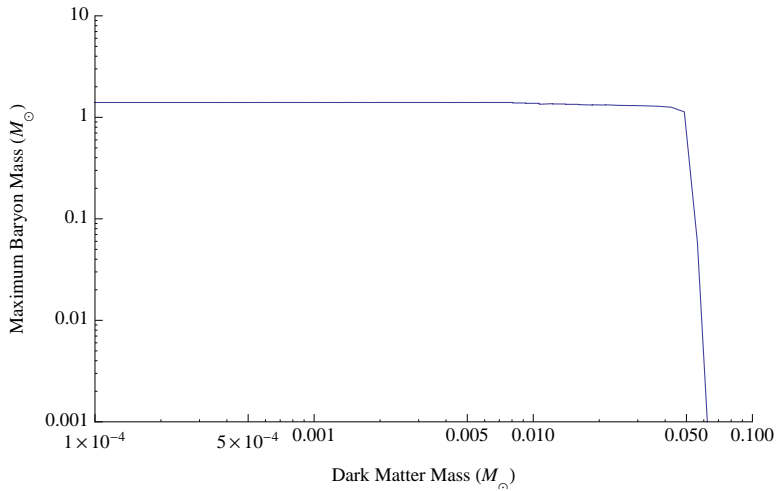}
   \caption{Relation Between Maximum Baryon Mass of White Dwarf and Dark Matter Core Mass}
   \label{fig4}
\end{figure}

Fig.\ref{fig4} shows the relation between maximum baryon mass and core mass. The increase of core mass reduces maximum baryon mass. There is a sharp decrease when the core mass exceeds $0.04M_{\odot}$. In this case the maximum baryon mass is not determined by stability consideration any more but the constraint on central density. When the core mass is even higher, over $0.06M_{\odot}$, there is no stable solution for white dwarfs. In this case the white dwarf most
likely collapses to neutron stars. We want to remark that although in our calculation we assume Fermi gas EOS for the entire white dwarf, realistically the surface boundary of a white dwarf is defined when the density go to zero. Obviously when density is sufficiently low the mass is no longer degenerate therefore the Fermi gas assumption is not valid. However the mass in the region where the Fermi gas approximation breaks down is very small therefore it does not affect our analysis.

\subsection{Neutron Stars}
In case of neutron stars, the mass of baryonic component inside the dark matter core is not negligible any more. We have to solve the two-fluid equations.
it should be noticed that in some cases, degenerate dark matter radius can be larger than the baryonic mass radius. We will illustrate this case in Fig. 6.

\begin{figure}[htbp] %  figure placement: here, top, bottom, or page
   \centering
   \includegraphics[width=4in]{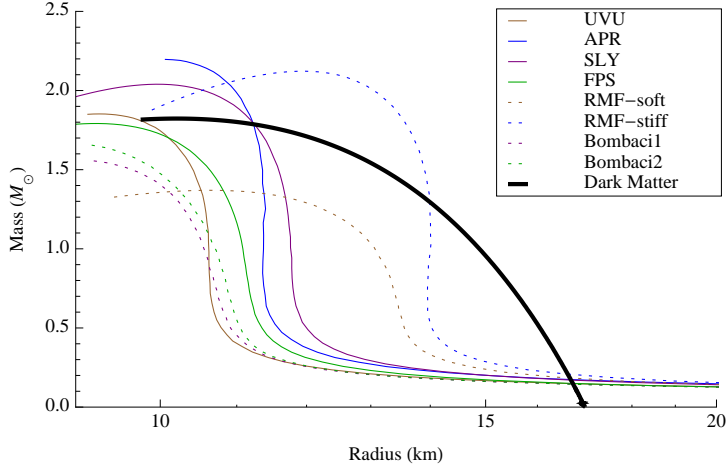}
   \caption{Mass-Radius Relation for Different EOS and Condensate Dark Matter}
   \label{fig5}
\end{figure}

We consider several popular types of EOS for the baryonic component, including UVU \cite{uvu}, APR \cite{apr}, SLy \cite{dh01}, FPS \cite{pr89}, RMF-soft and RMF-stiff \cite{kk97}, Bombaci1 and Bombaci2 \cite{bb97}. For the low density part of EOS, we adopt the BBPS \cite{bbps}. Fig. \ref{fig5} shows the Mass-Radius relation for these EOS's together with pure dark matter.

\begin{figure}[htbp] %  figure placement: here, top, bottom, or page
   \centering
   \includegraphics[width=4in]{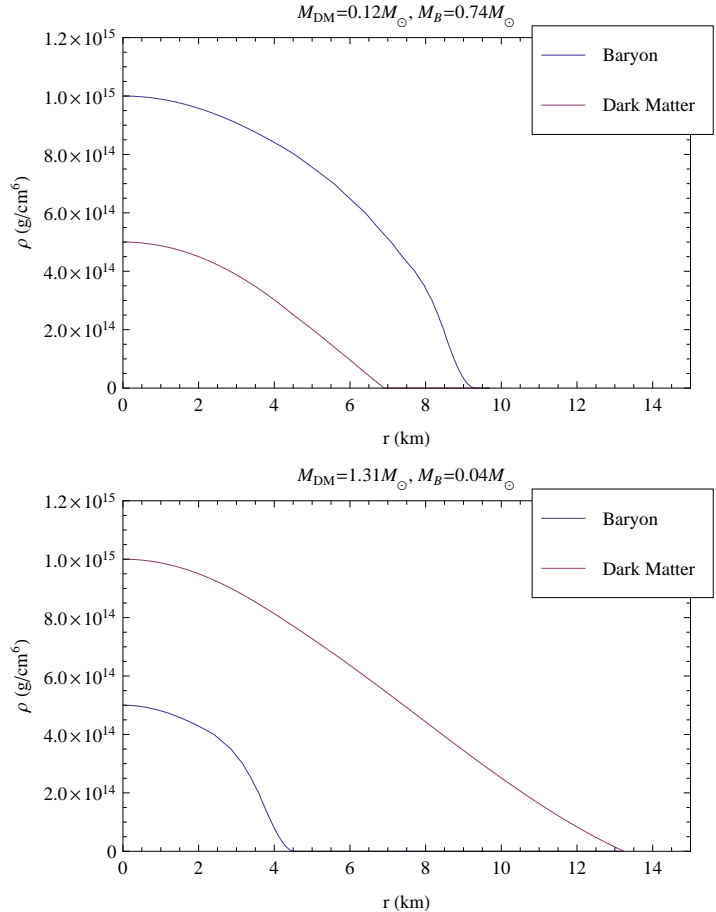}
   \caption{Examples of Structure for Two-fluid Neutron Stars}
   \label{fig6}
\end{figure}

The radius of such stellar object is determined when density becomes zero. The solution of two-fluid equations of generally gives different radius for baryonic and dark matter components. Fig.\ref{fig6}
gives two examples of stellar structure for baryonic and dark matter components using UVU EOS. We can see that dark matter radius can be
much larger than that of baryonic matter.
As we assume dark matter does not interact with normal matter except through gravity,
we take the radius when the baryonic density becomes zero as the radius for the two-fluid neutron star from the observational consideration,
while the observable mass is the total mass. This is very important.

\begin{figure}[htbp] %  figure placement: here, top, bottom, or page
   \centering
   \includegraphics[width=6in]{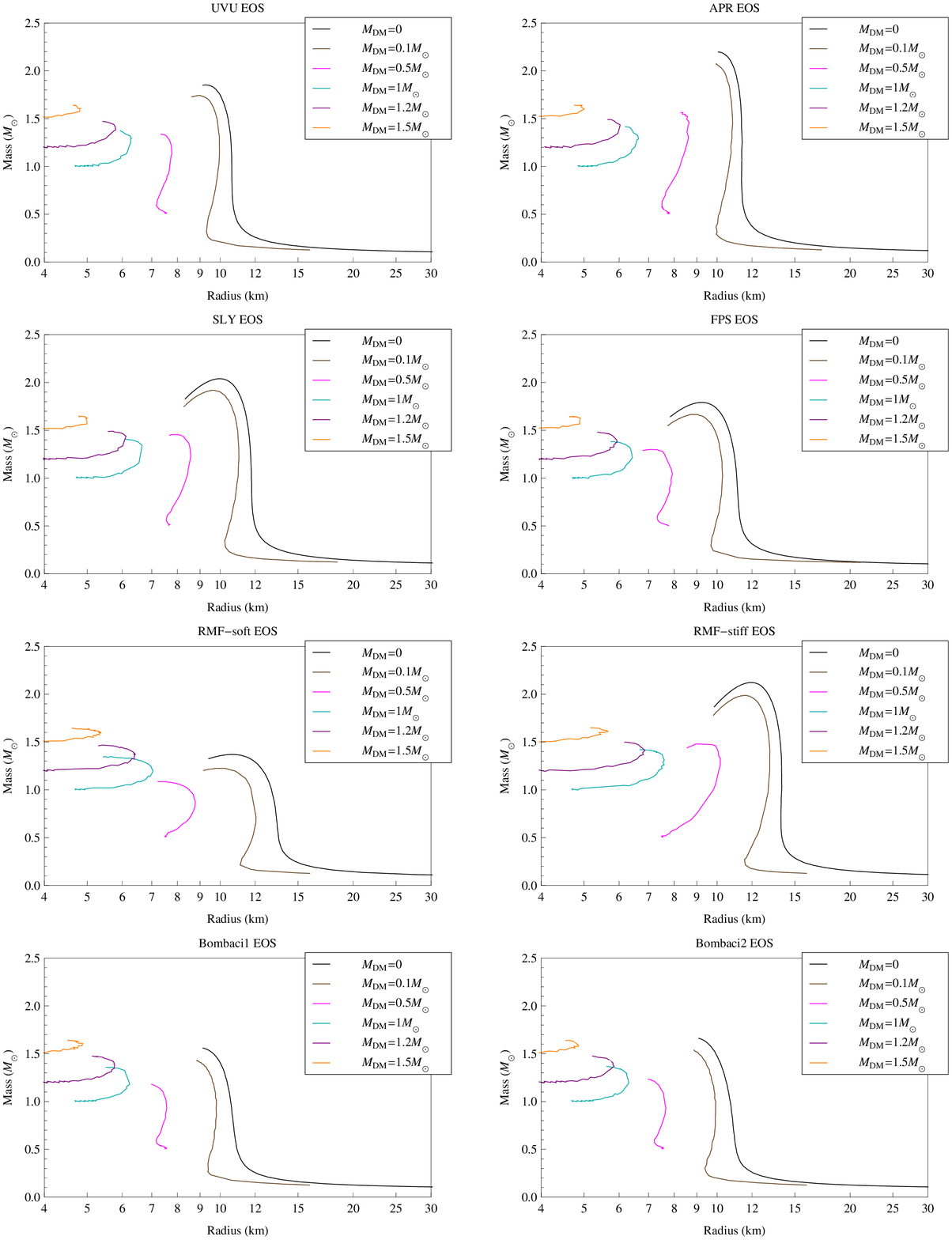}
   \caption{Mass-radius Relation  for Different EOS With Various Dark Matter Masses}
   \label{fig7}
\end{figure}

Fig.\ref{fig7} illustrates the relation between total mass and baryonic mass radius with the dark matter mass fixed. When the dark matter mass is small, the increase of dark matter mass deforms the original mass-radius relation, reducing the mass, radius and maximum mass. When dark matter mass becomes dominant, the maximum total mass increases again. For some EOS (RMF-soft and Bombaci1), the maximum mass can be higher than that of pure baryonic neutron star. The extreme case of a dark matter rich two-fluid star is that it becomes pure condensate dark matter. Condensate dark matter star has a maximum mass $\sim1.8M_{\odot}$, so the maximum mass will finally reach this value. For dark matter rich two-fluid stars, the maximum total mass increases with dark matter mass and in some cases exceeds the maximum mass of pure baryonic neutron star.

\begin{figure}[htbp] %  figure placement: here, top, bottom, or page
   \centering
   \includegraphics[width=6in]{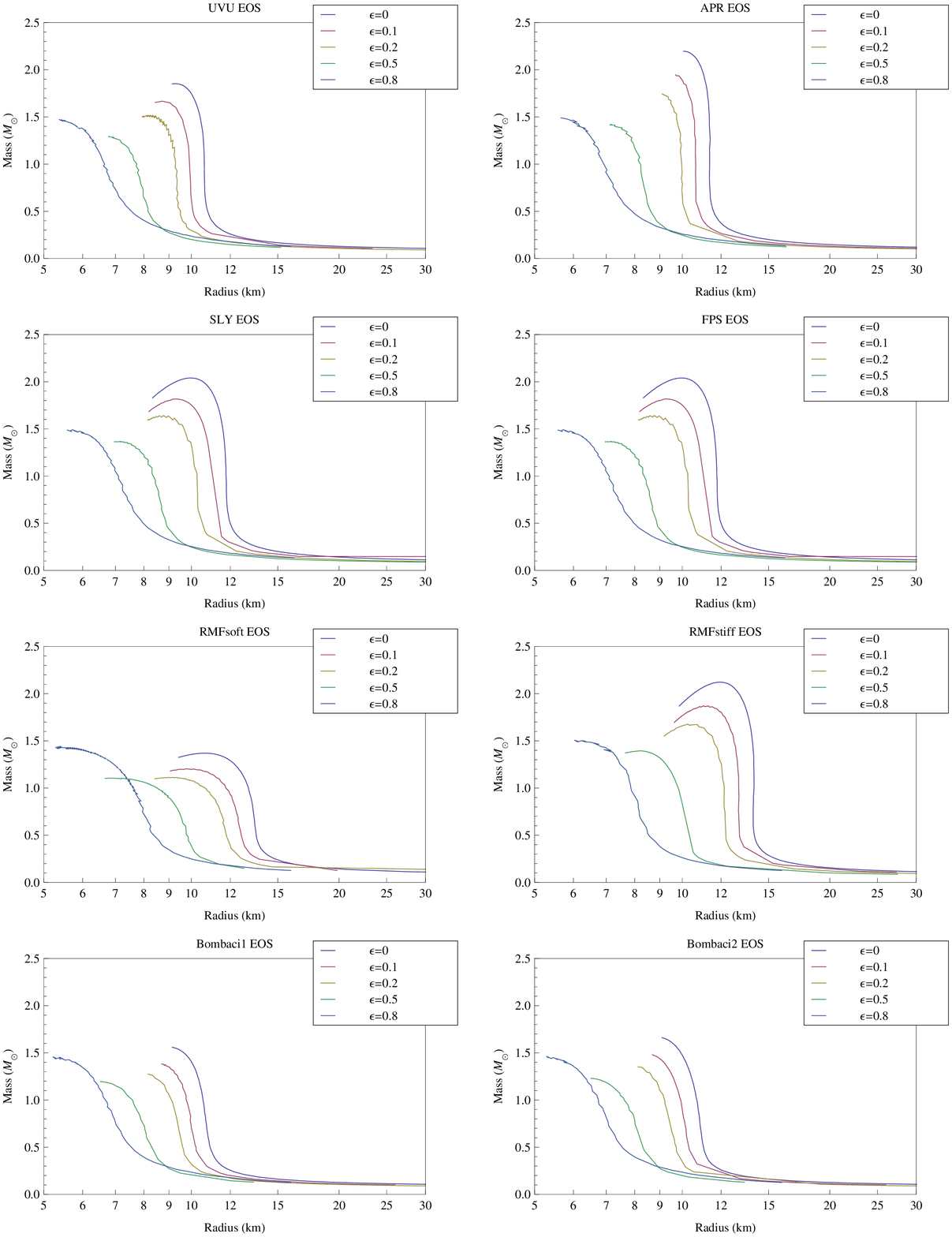}
   \caption{Mass-radius Relation  for Different EOS With Various Fixed $\epsilon=\frac{M_{\rm DM}}{M_{\rm B}+M_{\rm DM}}$}
   \label{fig8}
\end{figure}

Fig.\ref{fig8} shows the relation between total mass and radius with dark matter energy ratio $\epsilon$ fixed. The energy ratio $\epsilon$ is defined by $\epsilon=\frac{M_{\rm DM}}{M_{\rm B}+M_{\rm DM}}$. The increase of $\epsilon$ reduces the mass and radius while generally keeping the shape. When dark matter dominates, the shape transforms that of pure dark matter, but the radius is smaller.

\begin{figure}[htbp] %  figure placement: here, top, bottom, or page
   \centering
   \includegraphics[width=4in]{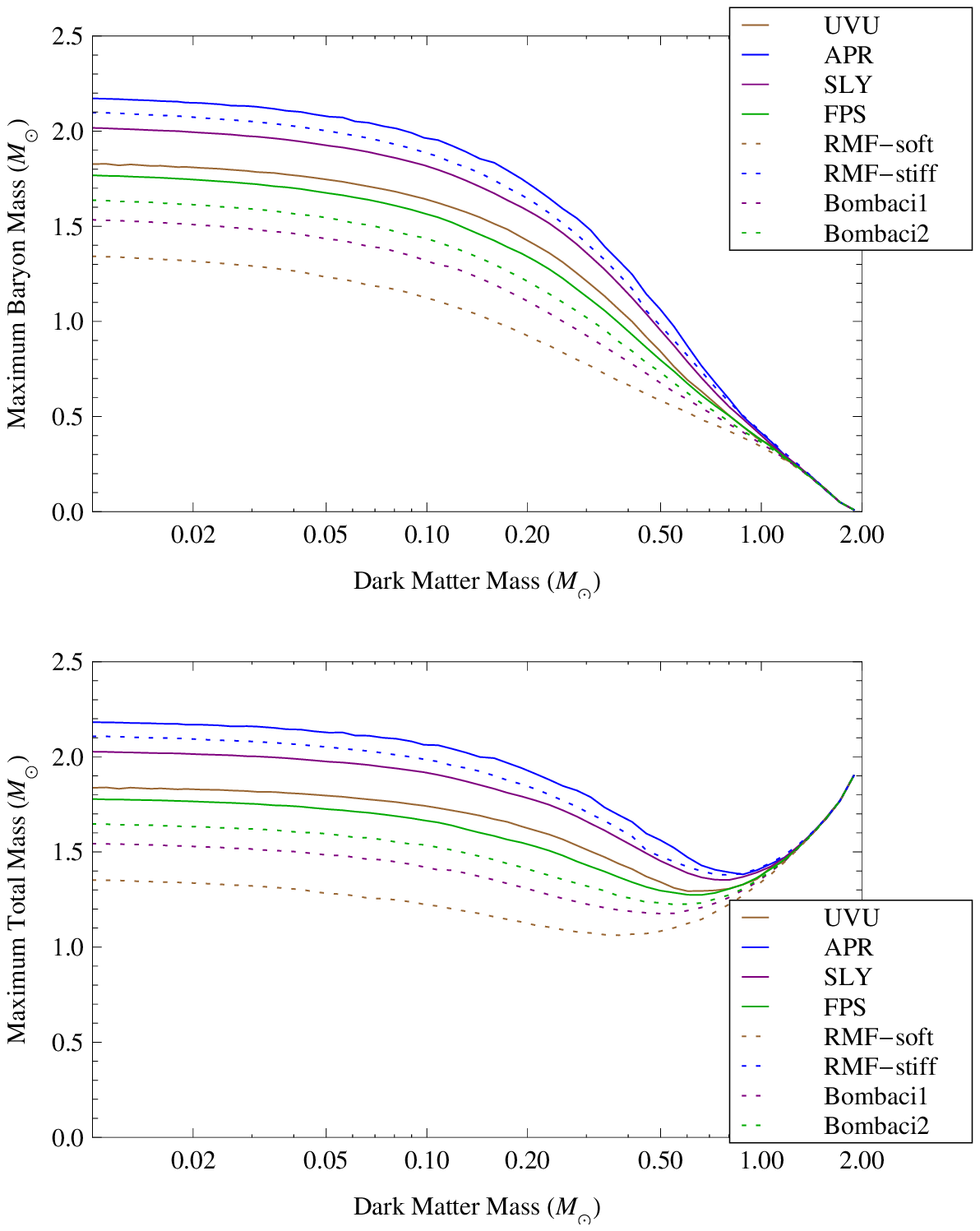}
   \caption{Relation  Between Maximum Mass of the Star and Dark Matter Mass. The upper panel is the maximum baryon mass vs dark matter and the lower panel is total mass (Baryon mass plus dark matter) vs dark matter.}
   \label{fig9}
\end{figure}

Fig.\ref{fig9} shows the relation between maximum mass of baryonic component  and maximum total mass with the dark matter mass. The increase of dark matter mass always reduces the maximum baryonic mass, but the maximum total mass increases after dark matter becomes dominant.

\section{Discussion}
The dark matter density outside the star is a key factor for accretion. After recombination, the dark matter density evolves proportional to $(1+z)^3$. In the early university the ambient density of dark matter is much higher and they can form high-density dark matter halos. Simulations reveal "cuspy" density profile of dark matter halos \cite{nfw96}. The density increases exponentially towards the halo center. Even though strong density cusps are not observationally favorable (cf. \cite{Kravtsov98}), the existence of deep gravitational well (massive stars or black holes) will pull the dark matter  and increase the density around them. We have calculated the increase of density using Blumenthal's method \cite{blu86} assuming only circular motion of dark matter particles. Freese et al. \cite{fr09} have calculated the change of density using a general treatment of adiabatic compression by Young \cite{Young80} considering the noncircular motion. The difference between the two methods are within a factor of 2 \cite{fr09}. Though adiabatic contraction formally requires the particles' orbital time to be shorter than the collapse time,  in practice the method works well beyond this limit \cite{st78}. The contracted dark matter density can be $\sim 10^{10}\rm GeV/cm^3$ at $10^{-4}$pc from the gravitational center \cite{fr09}. We hope that more knowledge shall be obtained by future simulations and observations of small scale structures. Furthermore although the average dark matter density in solar system is very low, it is possible that much higher density dark matter may exist near the galactic center, e.g. \cite{abdo10a, abdo10b}, so stars in this region may accrete enough dark matter.

By assuming the dark matter density is sufficiently high around the stars,e.g. in the dark matter halos of the early universe or at the center of the galaxies, we study the gravitational effect of condensate dark matter on various stellar objects. The deep gravitational well created by the stars pulls dark matter onto it. The self-interaction of dark matter particles enables them to lose energy and rest inside the star. When the dark matter density exceeds the critical value, the phase transition to condensation state takes place. Condensate dark matter develops a compact core and alter the structure and stability of the stellar objects. Normal stars will become unstable except for extremely light dark matter core. For white dwarf, the maximum dark matter core is about $0.06M_{\odot}$. Condensate dark matter admixed neutron stars is studied through the two-fluid TOV equation. The existence of condensate dark matter has significant impact on the mass-radius relation and lower the maximum mass compared to neutron stars. Condensate dark matter admixed neutron star can result in compact objects with very small coordinate radius of visible object, which is determined by the baryonic matter.
The existence of degenerate dark matter will also lower the mass threshold for stars to
collapse to a black hole. This may lead to more GRBs caused by
stellar collapse. Recent studies show that the rate of GRBs
does not strictly follow the star formation history but may be actually
enhanced by some other mechanisms at high-redshift \cite{Kistler09,Wang09, Cheng10}.
The standard collapsar model indicates that stars with mass larger than
$30M_{\odot}$ can produce GRBs \cite{Woosley93}. The existence of
degenerate dark matter core can reduce the mass threshold of stars that can produce
GRBs. If $30M_{\odot}$ is the threshold for forming GRBs from stars,
the threshold may be down to $20M_{\odot}$ after including
gravitational effect of condensate dark matter effect. For a
Salpeter initial mass function of stars with lower and upper mass
cutoffs $m_{\rm l}=0.1M_\odot$ and $m_{\rm
u}=100M_\odot$\cite{Salpeter55}, the percentage of stars that can
die as black holes increases by 90\%. So the star mass threshold of
GRBs formation decreased due to dark matter effect could be an
alternative solution to the excess of the high-redshift GRB rate.

Leung, Chu and Lin \cite{lcl11} also studied the property of dark matter admixed neutron stars. We differ from them in several key respects. First, they assume dark matter as ideal Fermi gas but we study the Bose-Einstein condensate dark matter, which have very different EOSs. Second, Leung et al. \cite{lcl11} did not consider how dark matter gets into the star, what are the key factors to determine the mass of dark matter inside the star and how dark matter is possible to rest inside the neutron star. We consider the accretion process through dark matter self-interaction from the surrounding halo. Finally, Leung et al. \cite{lcl11} studied the two-fluid model from first principle calculation. Though the numerical results from their formulations and ours formulations given by \cite{sc09} and \cite{cs11} are identical \cite{chu}, their approach can be used to study the time-dependent properties, like stellar oscillations \cite{lcl12}.

We thank Dr. T. C. Harko and Prof. M.C. Chu for useful discussion and suggestions and the anonymous referee for his very helpful suggestions and comments.


\begin{thebibliography}{99}
\bibitem{st78}G. Steigman et al., AJ, 83, 1050S (1978)
\bibitem{sp85}D. N. Spergel, and W. H. Press, ApJ, 294, 663 (1985)
\bibitem{fg85}J. Faulkner and R. L. Gilliland, ApJ, 299, 994 (1985)
\bibitem{bf08}G. Bertone, M. Fairbairn, Phys. Rev. D, 77, 043515 (2008)
\bibitem{k08}C. Kouvaris. Phys. Rev. D, 77, 023006 (2008)
\bibitem{gn89} I. Goldman and S. Nussinov, Phys. Rev. D, 40, 3221 (1989)
\bibitem{gd90}A. Gould, B.T. Draine, R.W. Romani and S. Nussinov, Physics Letters B, 238, 337 (1990)
\bibitem{sp09} D. Spolyar et al., PRL, 100, 051101 (2009)
\bibitem{spap09}D. Spolyar et al., ApJ, 705, 1031 (2009)
\bibitem{rip10} E. Ripamonti et al., MNRAS, 406, 2605 (2010)
\bibitem{cl09}J. Casanellas and I. Lopes, ApJ, 705, 135 (2009)
\bibitem{cl11}J. Casanellas and I. Lopes, ApJ, 733, L51 (2011)
\bibitem{fs10}M. T. Frandsen and S. Sarkar, PRL, 105, 011301 (2010)
\bibitem{cg10}D. T. Cumberbatch, J. A. Guzik, J. Silk, L. S. Watson and S. M. West, Phys. Rev. D, 82, 103503 (2010)
\bibitem{ti10}M. Taoso, F. Iocco, G. Meynet, G. Bertone and P. Eggenberger, Phys. Rev. D, 82, 083509 (2010)
\bibitem{cs11} P. Ciarcelluti and F. Sandin, Phys. Lett. B 695, 19 (2011)
\bibitem{sc09}F. Sandin and P. Ciarcelluti, Astropart. Phys., 32, 278 (2009)
\bibitem{lcl11}S. C. Leung, M. C. Chu, and L. M. Lin, Phys. Rev. D, 84, 107301 (2011)
\bibitem{bh07}C. G. Bohmer and T. Harko, JCAP, 06, 025 (2007)
\bibitem{sin1994} S.J. Sin, PRD, 50, 3650 (1994)
\bibitem{overduin2004} J.M. Overduin and P.S. Wesson, Phys Repts, 402, 267 (2004)
\bibitem{hu2000} W. Hu, R. Barkana and A. Gruzinov, PRL, 85, 1158 (2000)
\bibitem{ss96}E. Seidel and W. Suen, Proceedings of the Seventh Marcel Grossman Meeting on recent developments in theoretical and experimental general relativity, gravitation, and relativistic field theories. Edited by Robert T. Jantzen, G. Mac Keiser, and Remo Ruffini, River Edge, New Jersey: World Scientific, (1996)
\bibitem{bs98}J. Balakrishna, E. Seidel and W. Suen, Phys. Rev. D, 58, 104004 (1998)
\bibitem{bc98}J. Balakrishna, G. L. Comer, E. Seidel, H. Shinkai and W. Suen, Numerical Astrophysics : Proceedings of the International Conference on Numerical Astrophysics 1998 (NAP98), Edited by Shoken M. Miyama, Kohji Tomisaka, and Tomoyuki Hanawa. Boston, Mass. : Kluwer Academic (1999)
\bibitem{ch2012} P. Chavanis and T. Harko, PRD, 86, 4011 (2012)
\bibitem{li12}X. Y. Li, T. Harko and K. S. Cheng, JCAP, 06, 001 (2012)
\bibitem{gou87} A. Gould, ApJ, 321, 571, (1987)
\bibitem{blu86} G.R. Blumenthal et al, ApJ, 301, 27 (1986)
\bibitem{nfw96}J. F, Navarro, C. S. Frenk and S. D. M. White, ApJ, 462, 563 (1996)
\bibitem{bt94}J. Binney and S. Tremaine, {\it Galactic Dynamics}, Princeton University Press, Princeton (1994)
\bibitem{bkt}A. Burkert, ApJ, 447, L25 (1995)
\bibitem{abel02}Abel et al., Science, 295, 93 (2002)
\bibitem{fr09}K. Freese, P. Gondolo, J. A. Sellwood, and D. Spolyar, ApJ, 693, 1563 (2009)
\bibitem{fm97}Fukushige and Makino, ApJ, 447, J9 (1997)
\bibitem{ber85}E. Bertschinger, ApJS, 58, 39B (1985)
\bibitem{wan00}B. D. Wandelt et al., arXiv:astro-ph/0006344v2 (2000)
\bibitem{carter}B. Carter, {\it Relativistic Fluid Dynamics}, Springer-Verlag, Heidelberg (1989)
\bibitem{chu}S. C. Leung, M.Phil. Thesis, Chinese University of Hong Kong (2012)
\bibitem{bwn}S. L. Shapiro and S. A. Teukolsky, {\it Blach Holes, White Dwarfs and Neutrons Stars}, Wiley-VCH, Weinheim (2004)
\bibitem{ta}T. Padmanabhan, {\it Theoretical Astrophysics: Stars and stellar systems, Volume 2}, Cambridge University Press, Cambridge (2001)
\bibitem{fang}L. Z. Fang and S. P. Xiang, Acta Physica Sinica, 31, 9 (1982)
\bibitem{bbps}G. Baym, H. Bethe and C. J. Pethick, Nuclear Physics A, 175, 225 (1971)
\bibitem{uvu}R. B. Wiringa and V. Fiks, Phys. Rev. C, 38, 1010 (1988)
\bibitem{apr} A. Akmal, V. R. Pandharipande, and D. G. Ravelhall, Phys. Rev. C 58, 1804 (1998)
\bibitem{dh01}F. Douchin, and P. Haensel, A\&A, 380, 151 (2001)
\bibitem{pr89} V. R. Pandharipande and D. G. Ravenhall, {\it Hot Nuclear Matter, in Nuclear Matter and Heavy Ion Collisions}, NATO ADS Ser., Vol. B205, ed. M. Soyeur, H. Flocard, B. Tamain, and M. Porneuf (Dordrecht: Reidel), 103 (1989)
\bibitem{kk97}S. Kubis and M. Kutschera, Phys. Lett. B, 399, 191 (1997)
\bibitem{bb97}M. Baldo, I. Bombaci, and G.F. Burgio, A\&A, 328, 274 (1997)
\bibitem{Kravtsov98}A. V. Kravtsov, A. A., Klypin, J. S., Bullock, \& J.R. Primack, ApJ, 502,48 (1998)
\bibitem{Young80}P. Young, ApJ, 242, 1232 (1980)
\bibitem{abdo10a}A.A. Abdo et al., JCAP, 4, 14 (2010a)
\bibitem{abdo10b}A.A. Abdo et al., PRL, 104, 1302 (2010b)
\bibitem{Kistler09}M. D. Kistler, et al. ApJ, 705, L104 (2009)
\bibitem{Wang09}F. Y. Wang \& Z. G. Dai, MNRAS, 400, L10 (2009)
\bibitem{Cheng10}K. S. Cheng, Yun-Wei Yu \& T. Harko, PRL, 104, 241102 (2010)
\bibitem{Woosley93}S. E. Woosley, ApJ, 405, 273 (1993)
\bibitem{Salpeter55}E. E. Salpeter, ApJ, 121, 161 (1955)
\bibitem{lcl12}S. C. Leung, M. C. Chu, and L. M. Lin, Phys. Rev. D, 85, 3528 (2012)
\end{thebibliography}
\end{document}